\documentclass[%
 reprint,
superscriptaddress,
nofootinbib,
 amsmath,amssymb,
 aps,
pra,
floatfix,
]{revtex4-2}

\usepackage{graphicx}% Include figure files
\usepackage{xcolor}
\usepackage{dcolumn}% Align table columns on decimal point
\usepackage{bm}% bold math
\usepackage{comment}
\usepackage{wrapfig}
\usepackage{natbib}
\usepackage{multirow}
\usepackage{makecell}
\usepackage{float}
\usepackage{titlesec}
\usepackage{units}
\usepackage{color}
\usepackage{url}

\usepackage[colorlinks]{hyperref}
\hypersetup{%
	plainpages=true,
	breaklinks=true,% not default in dvips mode, so we must specify
	hypertexnames=false,%not ideal, but needed when pagenums duplicate (`i' vs. `1')
	pageanchor=true,
	colorlinks=true,
	linkcolor={blue},
	citecolor={red},
	urlcolor={blue},
	anchorcolor={black}
}

\begin{document}

\preprint{APS/123-QED}

\title{Real-time detection of correlated quasiparticle tunneling events in a multi-qubit superconducting device}

\author{Simon Sundelin}
\thanks{These two authors contributed equally.\\
simsunde@chalmers.se, anlinus@chalmers.se}

\author{Linus Andersson}
\thanks{These two authors contributed equally.\\
simsunde@chalmers.se, anlinus@chalmers.se}

\author{Hampus Brunander}

\author{Simone Gasparinetti}
\thanks{simoneg@chalmers.se, https://202q-lab.se}

\affiliation{%
Department of Microtechnology and Nanoscience,
Chalmers University of Technology,
412 96 Gothenburg, Sweden
}

\date{\today}

\begin{abstract}
    Quasiparticle tunneling events are a source of decoherence and correlated errors in superconducting circuits. Understanding—and ultimately mitigating—these errors calls for real-time detection of quasiparticle tunneling events on individual devices.
    In this work, we simultaneously detect quasiparticle tunneling events in two co-housed, charge-sensitive transmons coupled to a common waveguide.
We measure background quasiparticle tunneling rates at the single-hertz level, with temporal resolution of tens of microseconds. Using time-tagged coincidence analysis, we show that individual events are uncorrelated across devices, whereas burst episodes occur about once per minute and are largely correlated. These bursts have a characteristic lifetime of 7~ms and induce a thousand-fold increase in the quasiparticle tunneling rate across both devices. In addition, we identify a rarer subset of bursts which are accompanied by a shift in the offset charge, at approximately one event per hour. Our results establish a practical and extensible method to identify quasiparticle bursts in superconducting circuits, as well as their correlations and spatial structure, advancing routes to suppress correlated errors in superconducting quantum processors.

\end{abstract}

\maketitle
\section*{Introduction}
% Intro to the problem with quasiparticle bursts
Characterizing the dynamics of quasiparticle (QP) tunneling in superconducting circuits is crucial to understand and mitigate errors in large quantum processors. This importance arises because QP tunneling across Josephson junctions is a significant source of decoherence in superconducting qubits~\cite{glazman_bogoliubov_2021, aumentado_nonequilibrium_2004, sun_measurements_2012, serniak_hot_2018}. Recent advances in material science have improved qubit coherence~\cite{tuokkola_methods_2025, bland_millisecond_2025}, at the same time as quantum processors continue to scale in size~\cite{acharya_quantum_2025}. This progress has increased the importance of understanding and mitigating the detrimental effects of QP tunneling, in particular bursts of QPs that create correlated errors across multiple qubits~\cite{mcewen_resolving_2022, mcewen_resisting_2024, kurilovich_correlated_2025}. One of the motivations behind these efforts is that quantum error-correction codes, such as the surface code~\cite{fowler_surface_2012}, rely on errors being temporally and spatially uncorrelated. While this assumption is generally valid for relaxation due to material defects and uncorrelated photon-caused QP tunneling, it does not hold for ionizing- or phonon-only-induced pair-breaking events.

% General about QP tunneling sources and some mitigation strategies

Physically, Cooper-pair breaking occurs when the absorbed energy exceeds the superconducting gap $\Delta$ set by the thin-film material. The energy can originate from the absorption of high-energy photons~\cite{houzet_photon-assisted_2019}, or from interaction with high-energy phonons in the substrate~\cite{yelton_correlated_2025}. With improved shielding and high-frequency filtering, QP tunneling rates can be suppressed by orders of magnitude in the same device~\cite{serniak_direct_2019, gordon_environmental_2022, rehammar_low-pass_2023, connolly_coexistence_2024, andersson_co-designed_2025}. However, such measures do not protect against ionizing or phonon-only events, which cause bursts of QPs and extended periods of degraded relaxation times in many qubits simultaneously~\cite{vepsalainen_impact_2020, cardani_reducing_2021, martinis_saving_2021, iaia_phonon_2022, mcewen_resolving_2022, xu_distributed_2022, thorbeck_two-level-system_2023, fowler_spectroscopic_2024, mcewen_resisting_2024, harrington_synchronous_2025, kurilovich_correlated_2025, larson_quasiparticle_2025, li_cosmic-ray-induced_2025, valenti_spatial_2025, yelton_correlated_2025, nho_recovery_2025}.

% Sources of QP bursts

These bursts of QPs seem to stem from ionizing cosmic radiation or radioactive materials close to the device~\cite{mcewen_resolving_2022, vepsalainen_impact_2020, wilen_correlated_2021, cardani_reducing_2021, harrington_synchronous_2025, li_cosmic-ray-induced_2025, martinis_saving_2021, valenti_spatial_2025, larson_quasiparticle_2025}, or from phonon stress release in the vicinity of the chip~\cite{yelton_correlated_2025}. A key signature of ionizing events is a distinct change in the charge landscape around the qubit, which can be detected as a change in the charge offset of the transmon~\cite{christensen_anomalous_2019, wilen_correlated_2021, larson_quasiparticle_2025}. The phonon-only events are not expected to change the charge in the environment, but still cause bursts of elevated QP tunneling rates~\cite{yelton_correlated_2025}.

% On-chip mitigation strategies

On the chip level, there are a number of things that can be done to enhance the qubits' resilience against QP tunneling. For instance, engineering the gap difference of the junction has been proven to suppress QP tunneling and increase resilience against bursts~\cite{aumentado_nonequilibrium_2004, mcewen_resisting_2024, nho_recovery_2025}. To increase the effectiveness of gap engineering, combining this technique with improved thermalization to the cryogenic environment could also help thermalize hot phonons and reduce the harmful effects of the bursts~\cite{martinis_saving_2021, iaia_phonon_2022, larson_quasiparticle_2025}.

% Techniques for measuring QP tunneling in transmon qubits

Real-time monitoring of QP tunneling in qubits can be done in multiple ways~\cite{riste_millisecond_2013, serniak_direct_2019, amin_direct_2024}. For an offset-charge-sensitive transmon coupled to a readout resonator, one can monitor the charge-parity states of the qubit by mapping them to the ground and excited states of the transmon using coherent pulses~\cite{riste_millisecond_2013}. Another method is to perform joint, direct dispersive readout of qubit states and charge parity, in which the charge sensitivity of higher excited states of the transmon enables distinguishability of the charge-parity states~\cite{serniak_direct_2019, connolly_coexistence_2024}. A third approach is to couple an offset-charge-sensitive transmon directly to a waveguide and monitor the coherent scattering to reveal the charge-parity states~\cite{amin_direct_2024, fink2024}.

% Our contribution

In this work, we use this last method to study correlated QP bursts between two moderately charge-sensitive transmon qubits coupled directly to a waveguide. By continuously monitoring the coherent microwave scattering of both devices, we measure the QP-induced parity switching in real time. Occasionally, we observe a thousand-fold increase in the QP tunneling, persisting on average for about 7~ms. Most such events show temporal correlation, whereas the overall background tunneling shows no such correlation. For a small subset of burst events ($\sim 1-2 \mathrm{~h}^{-1}$, corresponding to a few percent of all bursts), we observe a distinct change in the background charge potential, compatible with ionizing events seen in other experiments~\cite{larson_quasiparticle_2025,wilen_correlated_2021, yelton_correlated_2025, christensen_anomalous_2019}.

\begin{figure}
    \includegraphics[width=8.8cm]{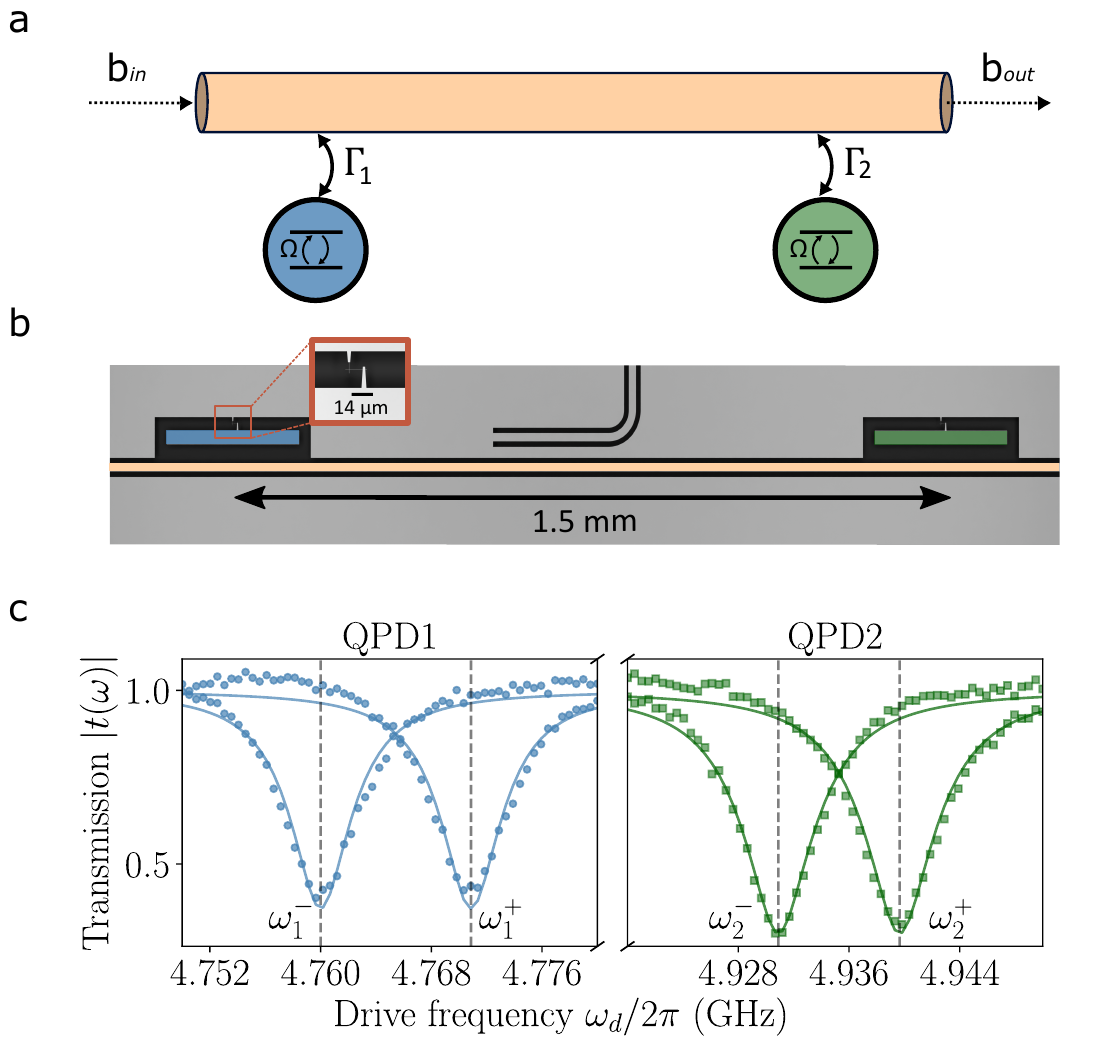}
    \caption{\label{fig1} \textbf{Device schematic and spectroscopy.} (a) Schematic of two transmon qubits (each approximated as a two-level system) coupled to a common coplanar waveguide with coupling rates $\Gamma_1$ and $\Gamma_2$. A coherent field $b_{\rm in}$ is sent through the waveguide, and the transmitted field $b_{\rm out}$, is measured at drive frequency $\omega_d$, defining the transmission coefficient $t = \frac{b_{\rm out}}{b_{\rm in}}$. (b) False-colored optical micrograph of the device, where the charge-sensitive transmons are highlighted in blue and green, and the coplanar waveguide is shown in light orange. (c) Representative magnitude of the transmission coefficient for the two detectors being in the even ($+$) or odd ($-$) charge-parity state. The corresponding transition frequencies $\omega_{1}^{\pm}$ and $\omega_{2}^{\pm}$ are indicated by dashed lines. Solid lines show fits to the theoretical transmission response obtained from master-equation simulations combined with input–output theory.}
\end{figure}

\begin{figure*}[t]
    \centering
    \includegraphics[width=18cm]{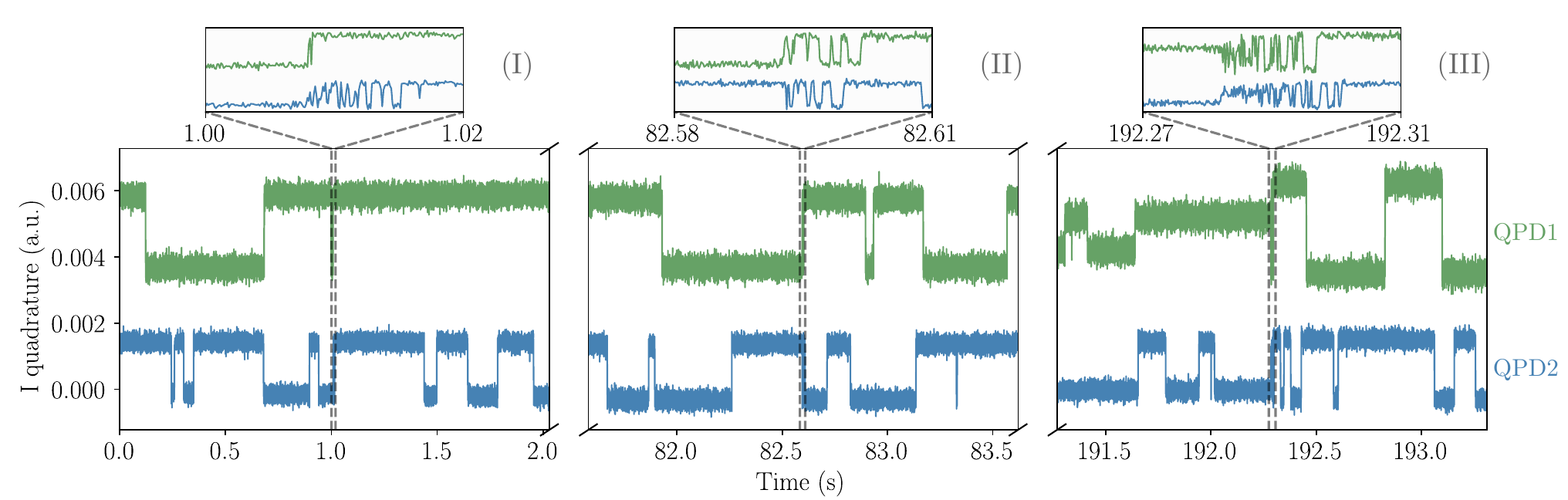}
    \caption{\label{fig2}\textbf{Representative real-time trajectories and burst classification.} Segmented time traces of the demodulated transmission signal for QP detector 1 (QPD1; top traces, green) and QP detector 2 (QPD2; bottom traces, blue), acquired during simultaneous continuous driving at the two parity-dependent transition frequencies $\omega_{01}^{\pm}$ for each device. Individual QP tunneling events are resolved as abrupt transitions between two discrete signal levels (low $\leftrightarrow$ high). In addition to the baseline switching dynamics, we observe intermittent intervals of strongly enhanced activity (“bursts”). The three panels illustrate the burst categories used throughout this work: an uncorrelated burst (I), where elevated switching occurs predominantly in one detector; a correlated burst (II), where both detectors exhibit increased switching over the same time window; and an offset-charge–shifting event (III), where a sudden change in the effective offset charge shifts $\omega_{01}^{\pm}$ and changes the parity-state discrimination. Top insets show expanded views of the time windows indicated by the dashed vertical markers.}
\end{figure*}

\section*{Results}
\subsection*{Detection of QP tunneling events}
Our device comprises two charge-sensitive, single-island transmon qubits coupled to a common coplanar waveguide [see Fig.~\ref{fig1}(a)]. In a transmon, the ratio between the Josephson energy $E_J$ and charging energy $E_C$ sets how strongly the qubit transition frequency $\omega_{01}$ depends on the offset charge $n_g$ (in units of Cooper pairs) on the superconducting islands \cite{koch_charge-insensitive_2007}. For our chosen ratios of $E_J/E_C\approx15$, the two transmons are moderately charge-sensitive. As a consequence, their fundamental transition frequencies follow two distinct branches associated with either even or odd electron-number parity states, $\mathcal{P}= \pm 1$. When a QP tunnels on or off an island through the Josephson junction, the charge parity flips, which is equivalent to shifting the offset charge by half a Cooper pair. This drives the qubit from one parity branch to the other and produces an instantaneous, resolvable shift of the transition frequencies, which we denote by $\omega_{1,2}^{\pm}$ for the two qubits. By monitoring such frequency shifts in real time, we determine when QP tunneling events take place \cite{amin_direct_2024}. In the following, we refer to our two transmons as QP detector one (QPD1) and two (QPD2).

We perform all measurements in a cryogenic setup~\cite{krinner_engineering_2019, gordon_environmental_2022}, with the chip mounted on the 10mK stage of a dilution refrigerator. Our device is fabricated on a silicon substrate with an aluminum ground plane and aluminum Josephson junctions. The two transmons share the same capacitor geometry and dimensions (island pads shown in green and blue in Fig.~\ref{fig1}b), but have different junction areas, and are separated by 1.5 mm along the coplanar waveguide. To suppress high-frequency radiation incident on the sample~\cite{connolly_coexistence_2024}, the chip is mounted in a sample holder inside an indium-sealed copper enclosure. In addition, all input and output lines are filtered using High-Energy Radiation Drain filters~\cite{rehammar_low-pass_2023, nho_recovery_2025, bland_millisecond_2025, andersson_co-designed_2025}.

To characterize QPD1 and QPD2, we send a coherent tone at frequency $\omega_d$ through the waveguide and measure the transmission coefficient. At resonance with the qubit transition,
%($\Delta = 0$)
and for weak enough driving, the coherently scattered field interferes destructively with the driving field, leading to a dip in the magnitude $|t(0)|$. In Fig.~\ref{fig1}(c), we show a representative trace of the averaged magnitude of the transmission coefficient for instances when the two detectors are in their even and odd parity states, respectively.
To each trace, we fit the model expression for the transmission coefficient of a two-level system coupled to a waveguide (see Supplemental Materials):
\begin{equation}
    t(\Delta) = \frac{2(i \Gamma \Delta + 2 \Delta^2 + \Omega^2)}{\Gamma^2 +4\Delta^2+ 2\Omega^2} \ ,
    \label{transmission}
\end{equation}
where $\Delta$ is the detuning between the drive and the qubit frequency, $\Gamma$ is the coupling strength between the qubit and waveguide, and $\Omega$ is the Rabi frequency, which is proportional to the drive amplitude.

From the parity-dependent frequency shifts, we extract maximum charge sensitivities of $\max _{n_g}^{(1)}=\left|\omega_{1}^{+}-\omega_{1}^{-}\right| / 2 \pi=11$ MHz for QPD1 and $\max _{n_g}^{(2)}=\left|\omega_{2}^{+}-\omega_{2}^{-}\right| / 2 \pi=9$ MHz for QPD2. Their respective average transition frequencies are $\bar{\omega}_{01}^1 / 2 \pi=4.766$ GHz and $\bar{\omega}_{01}^2 / 2 \pi=4.936$ GHz. We further extract the coupling rates to be $\Gamma_1/ 2 \pi=3.8$ MHz and $\Gamma_2/ 2 \pi=4.7$ MHz.

We carry out real-time monitoring of QP tunneling using the following measurement protocol. We first perform repeated single-tone spectroscopy of the transmitted magnitude, $|t(\omega)|$, for both devices. From this data we extract the parity-dependent transition frequencies $\omega_{01}^{\pm}$. To track parity switching in real time, we simultaneously apply two continuous drive tones at the measured transition frequencies $\omega_{01}^{\pm}$ (for both devices), record the transmitted field, digitally demodulate at both $\omega_{01}^{\pm}$ frequencies, and form a linear combination of the two demodulated traces to maximize the parity-state distinguishability \cite{amin_direct_2024}. The transmitted signal is also amplified using a traveling-wave parametric amplifier (TWPA) in the measurement chain to further improve the SNR. We acquire each time trace for 60 s with a bin size of 100 $\mu$s. Parity switches appear as abrupt transitions between two well-resolved signal levels (low $\leftrightarrow$ high), as illustrated by the representative traces in Fig.~\ref{fig2}. As seen in previous works \cite{amin_direct_2024}, we attribute these jumps to QP tunneling events onto or off of the superconducting island, which change the island charge parity and thereby shift the qubit transition frequency.

We repeat the spectroscopy and continuous-monitoring sequence over several hours; for each iteration, we re-estimate $\omega_{01}^{\pm}$ to compensate for slow drifts. Over the full experimental cycle, continuous monitoring constitutes 65\% of the time (QP-detection duty cycle), with the remaining 35\% spent on spectroscopy-based recalibration and acquisition overhead. The drifts are primarily driven by changes in the effective offset charge and can degrade the discrimination of switching events when the parity splitting $\left|\omega_{1}^{+}-\omega_{1}^{-}\right| / 2 \pi$ becomes comparable to, or smaller than, the relevant linewidth $\Gamma_{1,2}$. 

Apart from the baseline parity-switching dynamics, we observe intermittent “QP bursts”, during which the switching rate increases sharply. Out of these, we further classify the burst episodes into three categories (Fig.~\ref{fig2}). (i) Uncorrelated bursts: elevated switching activity is observed in one detector while the other remains at its baseline rate. (ii) Correlated bursts: both detectors exhibit a simultaneous increase in switching activity, consistent with temporally correlated QP tunneling events across the two devices. (iii) Offset-charge–shifting events: episodes accompanied by a detectable change in the effective offset charge, which shifts $\omega_{01}^{\pm}$ and changes the parity-state discrimination. We next analyze the statistics of these burst classes, focusing on tunneling rates, inter-device correlations, and associated offset-charge dynamics.

\subsection*{Burst statistics}
\begin{figure}
    \centering
    \includegraphics[width=8.8cm]{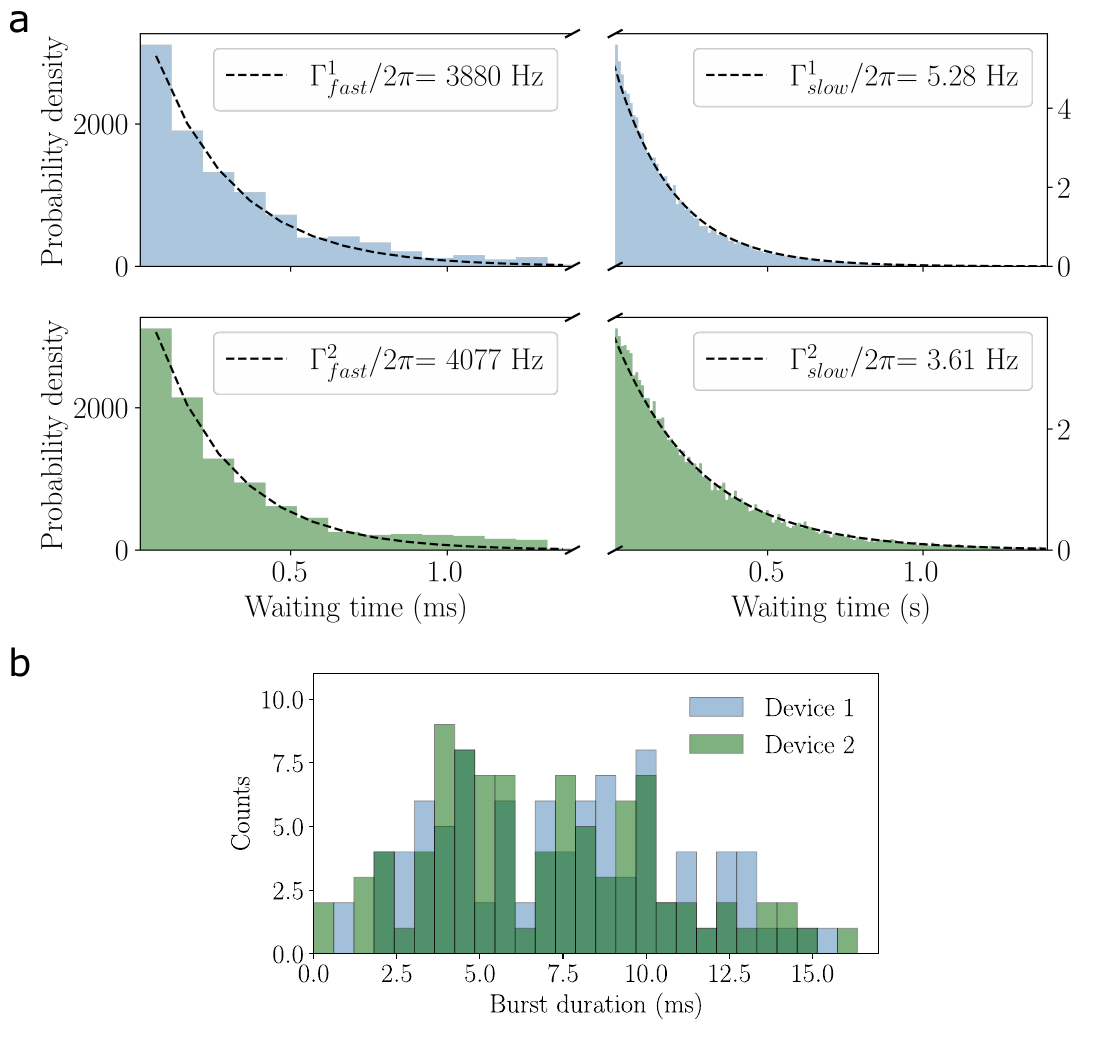}
    \caption{\textbf{Switching-rate extraction and burst-duration statistics.} (a) Histograms show the distribution of inter-event waiting times $t_{wait}$ between consecutive parity jumps detected by QPD1 (top, blue) and QPD2 (bottom, green), evaluated separately for burst intervals (left column, ms scale) and for the baseline switching dynamics (right column, s scale). Black dashed curves are single-exponential fits from which the extracted tunneling rates are acquired. (b) Burst-duration distribution. Histogram of burst durations $\tau_{burst}$ for QPD1 (blue) and QPD2 (green).}
    \label{fig3}
\end{figure}

We post-select time traces for which the parity-state discrimination fidelity exceeds $\mathcal{F} = 0.999$ (retaining $37\%$ of the acquired data) to avoid misidentifying the parity state when the parity splitting approaches the qubit linewidth. We then extract switching events from the demodulated time series using a simple thresholding protocol applied to the raw (otherwise unfiltered) quadrature streams; the only filtering is the effective low-pass response inherent to the demodulation and acquisition chain. Burst intervals are identified using kernel density estimation (KDE; see Supplemental Material) in combination with the criterion that bursts must be comprised of sequences of jumps with inter-event separations smaller than 5 ms. This procedure partitions the data into (i) baseline parity switching and (ii) burst-associated switching. For each detector, we construct the waiting-time distribution $P(t_{\rm wait})$ of the extracted events in each subset and fit a single-exponential model to it (Fig.~\ref{fig3}). This exponential decay is consistent with Poissonian switching within each regime. We obtain the corresponding tunneling rates from the mean waiting time $\Gamma =\frac{1}{\langle t_{wait} \rangle }$, yielding $\Gamma_{slow}^{(1)} = 5.28~\mathrm{s}^{-1}$ and $\Gamma_{slow}^{(2)} = 3.61~\mathrm{s}^{-1}$ for the baseline dynamics of detectors 1 and 2, respectively. Applying the same analysis within burst intervals gives $\Gamma_{fast}^{(1)} = 3880~\mathrm{s}^{-1}$ and $\Gamma_{fast}^{(2)} = 4077~\mathrm{s}^{-1}$, an enhancement by approximately three orders of magnitude relative to the background. Because the time traces are binned at 100 $\mu$s, we interpret these burst rates as lower bounds on the intrinsic tunneling rate during the fastest episodes. Finally, we attribute the observed difference in $\Gamma_{slow}$ between QPD1 and QPD2 to state-dependent QP dynamics in transmons \cite{connolly_coexistence_2024} and the detector-to-detector variations in the steady-state qubit population set by the applied drive conditions \cite{amin_direct_2024}.

Additional timescales that characterize the burst phenomenology are the burst occurrence rate and the burst duration. From the post-selected trajectories we extract the burst start and end times (Methods; Supplemental Material) and find that bursts in QPD1 and QPD2 occur at a rate of 1.6 min$^{-1}$ and 1.0 min$^{-1}$ respectively. This occurrence rate is consistent with prior observations of intermittent QP activity in superconducting circuits \cite{kurilovich_correlated_2025, nho_recovery_2025, diamond_distinguishing_2022, yelton_correlated_2025}. We further determine the burst duration distribution and obtain a mean duration of $\langle \tau_{burst}\rangle \approx 7$ ms for both detectors (Fig.~\ref{fig3}b). 

\begin{figure}
    \centering
    \includegraphics[width=8.8cm]{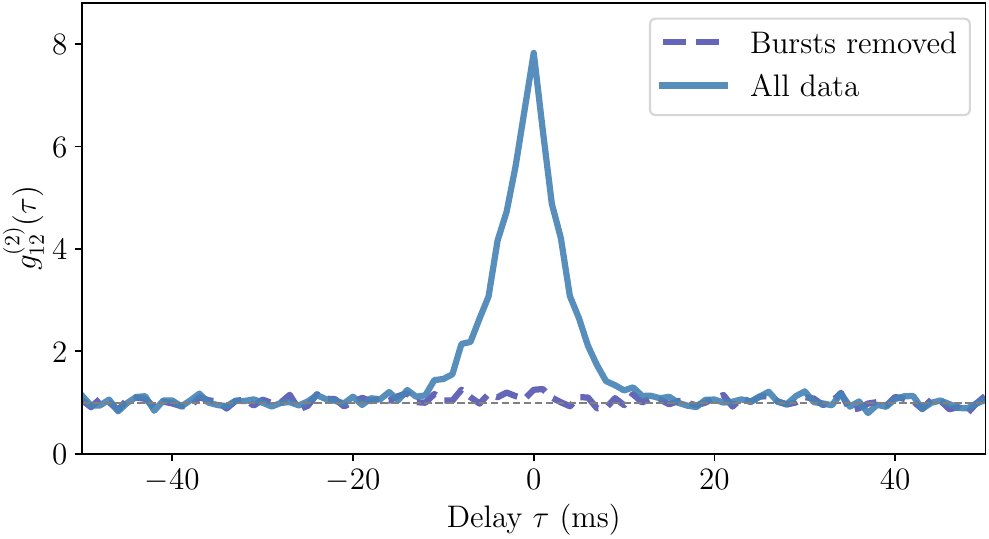}
    \caption{\textbf{Cross-correlation of burst activity between QP detectors.} Intensity cross-correlation $R_{12}(\tau)$ between QPD1 and QPD2 computed from the burst-count time series, where the intensity is defined as the number of identified burst events per $\Delta t=1~\mathrm{ms}$ bin. The solid trace (“All data”) shows a pronounced peak at $\tau=0$, $R_{12}(0)>1$, indicating temporally correlated burst activity across the two devices. At large $|\tau|$ the correlation approaches the uncorrelated baseline $R_{12}(\tau)\simeq 1$. The dashed trace (“Bursts removed”), obtained after excising burst intervals from both time series, yields $R_{12}(\tau)\approx 1$ for all $\tau$, consistent with independent Poissonian dynamics. }
    \label{fig4}
\end{figure}

We quantify inter-device correlations in the burst dynamics by computing the intensity cross-correlation $R_{12}(\tau)$ from the burst-count time series of the two detectors:
%Here the intensity $I_i(t)$ is defined as the number of identified burst events in detector $i\in{1,2}$ per non-overlapping time bin of width $\Delta t=1~\mathrm{ms}$. We evaluate
\begin{equation}
R_{12}(\tau)=\frac{\left\langle I_1(t) I_2(t+\tau)\right\rangle}{\left\langle I_1\right\rangle\left\langle I_2\right\rangle} \ ,
\end{equation}
where the intensity $I_i(t)$ is defined as the number of identified burst events in detector $i\in{1,2}$ per non-overlapping time bin of width $\Delta t=1~\mathrm{ms}$. We evaluate $R_{12}(\tau)$
over a delay range $|\tau|\le 50~\mathrm{ms}$, where $\langle\cdot\rangle$ denotes averaging over the segmented dataset. The resulting $R_{12}(\tau)$ exhibits a pronounced peak at zero delay, $R_{12}(0)>1$, demonstrating that burst activity is temporally correlated across the two devices (Fig.~\ref{fig4}). The peak is centered and approximately symmetric about $\tau=0$, indicating no measurable time ordering and hence no evidence for directional triggering between the two detectors within our temporal resolution. At larger delays, the correlation relaxes to a flat baseline $R_{12}(\tau)\approx 1$, consistent with uncorrelated switching outside burst intervals; correspondingly, after excising burst episodes from the time traces we obtain $R_{12}(\tau)\approx 1$ for all $\tau$, confirming that the background dynamics are well described as Poissonian and that the observed inter-device correlations originate from the burst events. We identify 197 burst episodes in QPD1 and 121 in QPD2. Of these, 92 occur coincidentally in both detectors (correlated bursts), corresponding to 
47\% of QPD1 bursts and 76\% of QPD2 bursts.

\subsection*{Offset-charge-shifting burst events}
\begin{figure}
    \centering
    \includegraphics[width=8.8cm]{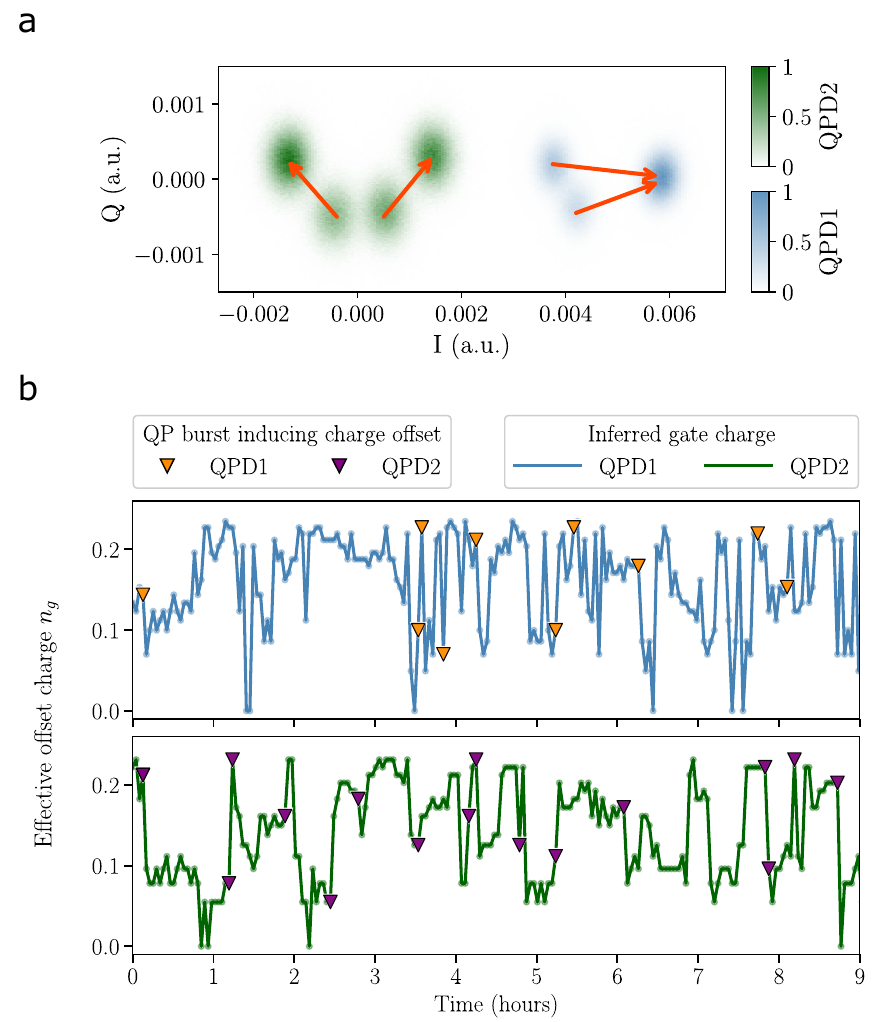}
    \caption{\textbf{Burst-associated offset-charge shifts in two QP detectors.} (a) Two-dimensional normalized histograms of the demodulated IQ response during continuous parity monitoring, showing the two well-resolved parity manifolds for each device. For offset-charge--shifting burst events, the cluster centroids exhibit a sudden displacement (red arrows), consistent with detuning of the monitoring tones due to a change in the effective offset charge. (b) Time evolution of the effective offset charge $n_g$ inferred from the parity splitting extracted from single-tone spectroscopy for QPD1 (top, blue trace) and QPD2 (bottom, green trace). Triangles indicate offset-charge--shifting burst episodes identified in the subsequent continuous QP-monitoring record (QPD1: orange; QPD2: purple); the leftmost markers correspond to a coincident offset-charge--shifting burst in both devices which results in the IQ-cluster displacements shown in panel (a). These events coincide with step-like changes in $n_g$, identifying a rare subclass of bursts associated with discrete offset-charge rearrangements.}
    \label{fig5}
\end{figure}

In addition to uncorrelated and correlated burst activity, we observe a third class of events in which a \emph{correlated burst event} is accompanied by an abrupt change in the parity-state discrimination. Operationally, these events appear as a sudden displacement of the two blobs in the demodulated IQ plane corresponding to the two parity states (Fig~\ref{fig5}a): the center of the cluster shifts, resulting in the applied monitoring tones no longer optimally align with the transition frequencies $\omega_{01}^{\pm}$. We classify such episodes as \emph{offset-charge-shifting bursts}, because the observed signature is consistent with a discrete change in the effective offset charge that modifies the parity splitting and therefore the state discrimination. 

We relate these discrimination-changing bursts to the independently measured evolution of the parity splitting. Specifically, before each continuous QP-monitoring segment, we perform a short spectroscopy measurement to re-extract the parity-dependent transition frequencies $\omega_{01}^{\pm}$ (as previously described), from which we compute the parity splitting $\Delta\omega_{01}^{\pm}/2\pi \equiv (\omega_{01}^{+}-\omega_{01}^{-})/2\pi$ and infer the corresponding effective offset charge $n_g$ (blue and green markers in the representative segments shown in Fig.~\ref{fig5}b; see Supplemental Material for details). The resulting time series of $n_g$ exhibits slow fluctuations punctuated by larger discrete steps, consistent with intermittent rearrangements of the effective offset charge. The triangular markers denote the offset-charge-shifting burst episodes that are identified in the subsequent continuous QP-monitoring record. Analyzing the full dataset (including intervals with $\mathcal{F}<0.999$) over a total acquisition time of 14 hours, we identify 14 charge-offset–shifting QP tunneling bursts in QPD1 and 22 in QPD2. Of these events, 6 exhibit simultaneous charge-offset jumps in both detectors. These counts correspond to rates of $1.4~\mathrm{h}^{-1}$ and $2.0~\mathrm{h}^{-1}$ for QPD1 and QPD2, respectively, and $0.43~\mathrm{h}^{-1}$ for coincident offset-charge–shifting bursts. Notably, these rates are much lower than the offset-charge--preserving  burst rates reported above, underscoring that offset-charge–shifting bursts constitute a rare subclass.

Moreover, offset-charge-shifting bursts for QPD1 (QPD2) account for $22\%$ ($57\%$) of all instances in which the measured change in $n_g$ exhibits a discrete step exceeding $\left|n_{g, i+1}-n_{g, i}\right| > 0.1$, where $n_{g, i}$ denotes the inferred effective charge offset extracted from the $i$th spectroscopy calibration. The remaining large steps are not accompanied by an observed offset-charge--shifting burst in the QP monitoring trace, likely for three reasons. First, the QP-detection duty cycle is 65\%, while the remaining 35\% of each experimental cycle is spent on spectroscopy and acquisition overhead, during which bursts cannot be directly recorded. Under the assumption that such events are uniformly distributed in time and are detected with unit efficiency when monitoring is active, this would correspond to an upper bound of $\sim 0.22/0.65 \approx 35\%$ ($0.57/0.65 \approx 88\%$) for QPD1 (QPD2) for the true coincidence rate. Second, the parity splitting depends periodically on the effective offset charge $n_g$. Consequently, an offset-charge jump can move the system to a point on this periodic dependence where the resulting change in $n_g$ (and hence the parity-state discrimination at the fixed monitoring frequencies) is small, producing little or no observable signature, even though a charge rearrangement could have occurred. Finally,  offset-charge jumps may originate from physical mechanisms distinct from those responsible for QP tunneling bursts and therefore need not be accompanied by enhanced QP activity.

In addition to these large discrete steps, we observe slow variations in $n_g$, consistent with a gradually evolving electrostatic environment. The observed coincidence with the larger step-like changes supports the interpretation that a subset of bursts is associated with abrupt offset-charge shifts. The preferential association between large step-like changes and burst activity is consistent with an energy-deposition event that both generates QP-tunneling and perturbs the local electrostatic environment. Such coincident QP and charge-offset signatures are commonly attributed to ionizing events (e.g., cosmic-ray), although we do not directly discriminate the underlying microscopic origin here.

\section*{Discussion}
We demonstrate real-time detection of correlated quasiparticle (QP) tunneling bursts using two moderately charge-sensitive transmon qubits directly coupled to a common waveguide. By continuously monitoring their charge-parity states via coherent microwave scattering, we resolve individual parity-switching events with a time resolution of 100~$\mu$s and identify intermittent burst episodes in which the QP tunneling rate increases by approximately three orders of magnitude.

A key result of this work is the identification of three distinct regimes of QP dynamics. Outside burst intervals, parity-switching events are well described by independent Poissonian processes and show no measurable temporal correlation between the two devices. During burst intervals, we observe both uncorrelated and correlated activity. In uncorrelated bursts, elevated switching rates appear in only one detector, whereas in correlated bursts both detectors exhibit a simultaneous increase in switching activity, manifested as a pronounced zero-delay peak in the second-order intensity cross-correlation.

Within the class of correlated bursts, we identify an additional subset of events that are accompanied by abrupt changes in parity-state discrimination and discrete shifts in the measured parity splitting, consistent with sudden rearrangements of the local charge environment expected from ionizing radiation events. The observation of correlated bursts both with and without detectable offset-charge shifts indicates that multiple physical mechanisms contribute to the observed QP dynamics, including ionizing radiation as well as phonon-mediated pair-breaking processes that do not directly perturb the electrostatic environment. 

One possible interpretation is that an ionizing event generates high-energy phonons that propagate over macroscopic distances and induce correlated QP bursts across multiple devices, while the associated rearrangement of charge remains localized near the interaction site, which would be consistent with results in previous experiments~\cite{yelton_correlated_2025,larson_quasiparticle_2025}. In this scenario, a correlated burst may originate from an ionizing event even if no offset-charge shift is detected in a given detector.
Corroborating this interpretation would require spatially resolved measurements beyond what we presented in this work.

Integrating multiple detectors on a single chip will enhance the spatial resolution of the burst-detection technique, similar to recent work utilizing an array of high-kinetic-inductance resonators~\cite{valenti_spatial_2025}. In this respect, we note that the analog bandwidth of our detectors is limited by the qubit linewidth, which can be made to be up to tens of MHz, corresponding to a temporal resolution down to a few ns. By performing edge detection on noisy single-shot traces and collecting statistics of the measured time differences, we anticipate being able to resolve delays well below the single-shot temporal resolution of our detectors, which is limited by sensitivity and added noise.

Further improvements to the sensitivity of our detectors could be achieved by reading out the detectors in reflection, in which all scattered radiation is coupled to a single measurement port, thereby increasing the effective signal-to-noise ratio. Integrating voltage control of the offset charge and operating the sensor at maximum parity state separation would increase the amount of useful data during the measurement period. The control scheme of the voltage could also be used to detect the charge-offset jumps. The duty cycle of the measurement sequence could also be improved by reducing the real-time processing latency, for example, by implementing binary thresholding of the readout signal.

Because our approach is based on superconducting qubits, it naturally enables the exploration of mitigation strategies specific to superconducting qubit architectures, such as gap engineering and phonon-trapping techniques.
Thanks to its ability to detect individual tunneling events, we expect our method to provide a substantially more sensitive benchmark for these strategies than existing approaches that rely on transient coherence-time degradation.

In summary, real-time, parity-sensitive scattering measurements offer a powerful framework for diagnosing—and ultimately mitigating—correlated error mechanisms that pose a central challenge for fault-tolerant superconducting quantum computation.

\begin{acknowledgments}
The device used in this work was fabricated in Myfab Chalmers, a micro and nanofabrication laboratory.
We are thankful to VTT Technical Research Centre of Finland for providing the TWPA used in this work.
This work was supported by the Knut and Alice Wallenberg foundation via the Wallenberg
Centre for Quantum Technology (WACQT). S.~Gasparinetti acknowledges
funding from the European Research Council via Grant No. 101041744 ESQuAT.
S.~Sundelin acknowledges funding from the European Union via Grant No.~101080167 ASPECTS. L. Andersson acknowledges funding from the European Union via Grant No.~101135240 JOGATE.

\end{acknowledgments}

%\section*{Author contributions}

\bibliographystyle{naturemag}
%\bibliographystyle{apsrev4-2}
%\bibliography{References_fromZotero_QPD-Bursts}
%\bibliography{References_fromZotero}
%apsrev4-2.bst 2019-01-14 (MD) hand-edited version of apsrev4-1.bst
%Control: key (0)
%Control: author (72) initials jnrlst
%Control: editor formatted (1) identically to author
%Control: production of article title (-1) disabled
%Control: page (0) single
%Control: year (1) truncated
%Control: production of eprint (0) enabled

\pagebreak
\widetext
\newpage
\begin{center}
\textbf{\large Supplemental Materials: Quantum refrigeration powered by noise in a superconducting circuit}
\end{center}
%%%%%%%%%% Merge with supplemental materials %%%%%%%%%%
%%%%%%%%%% Prefix a "S" to all equations, figures, tables and reset the counter %%%%%%%%%%
\setcounter{equation}{0}
\setcounter{figure}{0}
\setcounter{table}{0}
\makeatletter
\renewcommand{\theequation}{S\arabic{equation}}
\renewcommand{\thefigure}{S\arabic{figure}}
\renewcommand{\bibnumfmt}[1]{[S#1]}

\section{Experimental setup}

We present a wiring diagram of our experiment in Fig.~1S. To distinguish the bursts of quasiparticle tunneling from the background tunneling rate caused by high frequency radiation, it is important to reduce the number of incident photons on the chip. To achieve a light-tight shielding of our chip, we package it into a sample holder, carefully designed to minimize line-of-sight 
\begin{figure*}[h]
    \includegraphics[width=16.8cm]{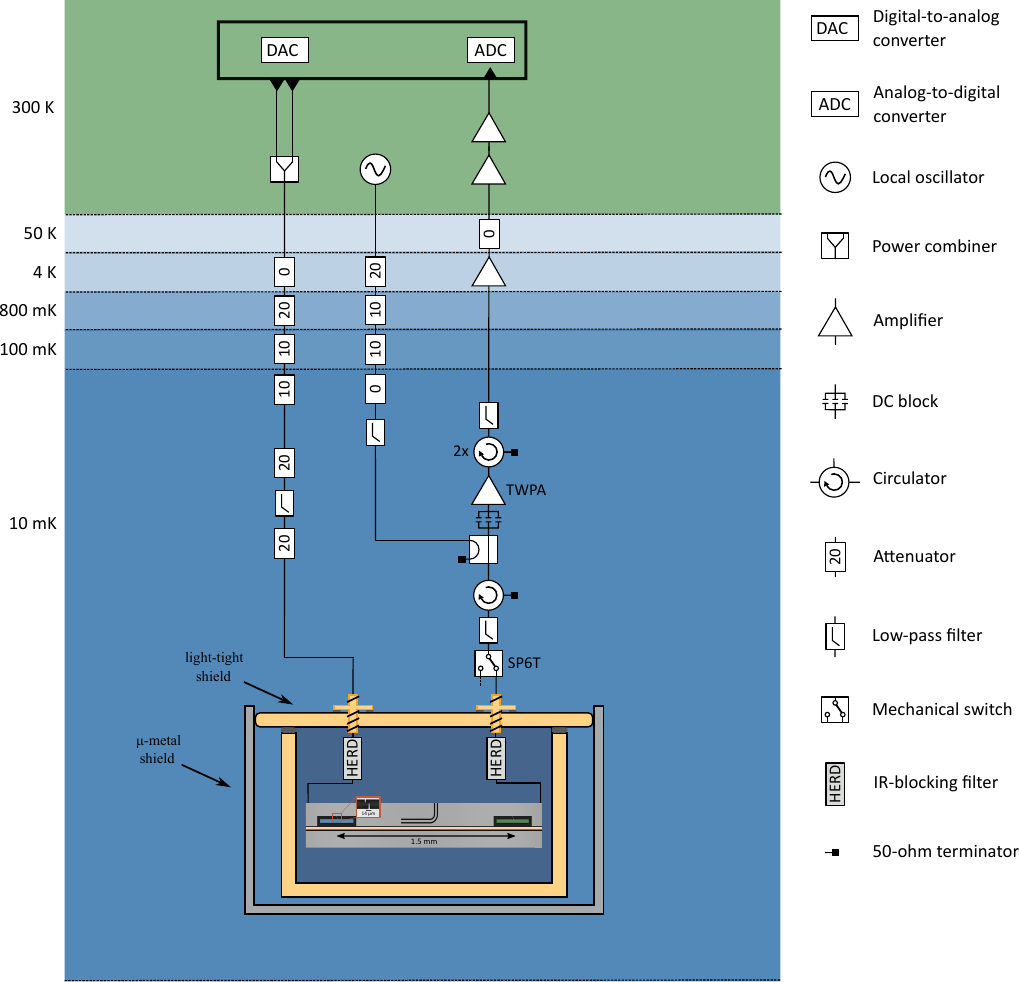}
    \caption{\label{fig:wiring} Experimental setup, see text for description.}
\end{figure*}radiation. Following best practices~\cite{connolly_coexistence_2024}, we employ an external copper enclosure in which all interfaces between adjacent parts, including the threaded SMA connectors, are sealed with indium O-rings to improve infrared tightness. To suppress infrared radiation in the coaxial lines, we employ a pair of High-Energy Radiation Drain (HERD) low-pass filters, which provide low in-band loss while simultaneously attenuating frequencies capable of breaking Cooper pairs~\cite{rehammar_low-pass_2023, andersson_co-designed_2025, nho_recovery_2025}. To further improve the signal-to-noise ratio of our readout chain, we use a Traveling Wave Parametric Amplifier (TWPA)~\cite{perelshtein2022}.

\section{\label{app:B} Spectroscopy model}
We model the qubit as a driven two-level system coupled to a waveguide with a coupling rate $\Gamma$. For a bidirectional waveguide we introduce right- and left-propagating coherent field amplitudes $\alpha_{\text {in}/\text{out}, R}$ and $\alpha_{\text {in} / \text {out}, L}$. For symmetric coupling into the two propagation directions, the directional decay rates satisfy $\Gamma_R=\Gamma_L=\frac{\Gamma}{2}.$ The corresponding input-output relations are 
$$
\alpha_{\mathrm{out}, R}=\alpha_{\mathrm{in}, R}-i \sqrt{\Gamma_R}\left\langle\sigma_{-}\right\rangle, \quad \alpha_{\mathrm{out}, L}=\alpha_{\mathrm{in}, L}-i \sqrt{\Gamma_L}\left\langle\sigma_{-}\right\rangle
$$
Driving from the left only corresponds to $\alpha_{\mathrm{in}, R} \equiv \alpha_{\mathrm{in}}$ and $\alpha_{\mathrm{in}, L}=0$. With the coherent input amplitude related to the Rabi frequency $\Omega$ as $\alpha_{\mathrm{in}}=\frac{\Omega}{\sqrt{2\Gamma}}$, the transmission coefficient reads

\begin{equation}
t \equiv \frac{\alpha_{\mathrm{out}, R}}{\alpha_{\mathrm{in}}}=1-i \frac{\Gamma}{\Omega}\left\langle\sigma_{-}\right\rangle .
\label{trans1}
\end{equation}
In the rotating frame of the drive, the qubit Hamiltonian is
$$
H=\Delta|e\rangle\langle e|+\frac{\Omega}{2}\left(\sigma_{+}+\sigma_{-}\right), \quad \Delta=\omega_p-\omega_{01}
$$
with $\sigma_{-}=|g\rangle\langle e|, \sigma_{+}=|e\rangle\langle g|$. At steady state, the density matrix obeys the Lindblad master equation
\begin{equation}     
\dot{\rho}=-i[H, \rho]+\Gamma \mathcal{D}\left[\sigma_{-}\right] \rho+\Gamma_{\mathrm{nr}} \mathcal{D}\left[\sigma_{-}\right] \rho+\frac{\gamma_\phi}{2} \mathcal{D}\left[\sigma_z\right] \rho = 0
\label{sME}
\end{equation}
where $\Gamma_{nr}$ accounts for additional decay channels other than the waveguide and $\gamma_{\phi}$ is the pure dephasing rate. The Lindblad superoperator is  
$$   
\mathcal{D}[c] \rho = c \rho c^{\dagger}-\frac{1}{2} (c^{\dagger} c \rho- \rho c^{\dagger} c).
$$
To evaluate $\left\langle\sigma_{-}\right\rangle$, we obtain the steady-state density matrix $\rho_{\mathrm{ss}}$ from eq \ref{sME} and compute $\left\langle\sigma_{-}\right\rangle=\operatorname{Tr}\left(|g\rangle\langle e|\rho_{\mathrm{ss}}\right)$. Combining with eq \ref{trans1} we get
\begin{equation}
t(\Delta)=1-\frac{i \Gamma \Gamma_{1}\left(\Delta-i \Gamma_{2 }\right)}{2\Omega^2 \Gamma_{2}+2\Gamma_{1 }(\Delta^2+\Gamma_{2 }^2)}.
\end{equation}
where $\Gamma_1 = \Gamma + \Gamma_{nr}$ and $\Gamma_2 = \frac{\Gamma_1}{2} + \gamma_{\varphi}$. 
Additionally, by letting $\Gamma_{nr} = 0$ and $\gamma_{\varphi} = 0$, we retrieve Eq (1) in the main text;
\begin{equation}
    t(\Delta) = \frac{2(i \Gamma \Delta + 2 \Delta^2 + \Omega^2)}{\Gamma^2 +4\Delta^2+ 2\Omega^2}.
    \label{transmission}
\end{equation}

\section{\label{app:C} Extraction of parity-switching events}
We continuously drive each QP detector at the two parity-dependent transition frequencies, $\omega_{01}^{+}$ and $\omega_{01}^{-}$, determined from the preceding spectroscopy calibration. A quasiparticle tunneling event onto or off of the superconducting island switches the charge parity, which shifts the effective transition frequency between $\omega_{01}^{+}$ and $\omega_{01}^{-}$. Because the device response at the fixed drive tones depends on the detuning from the instantaneous transition frequency, the demodulated complex envelopes of the transmitted field exhibit a two-level (telegraph) signal: when the system occupies a given parity, one tone is closer to resonance (larger response) while the other is farther detuned (smaller response), and the roles interchange upon a parity switch. This enables direct time-domain extraction of parity-switching (QP tunneling) events from the demodulated data streams.

For each device, we record two simultaneously demodulated complex time series, $s_{+}(t)$ and $s_{-}(t)$, corresponding to the two drive tones at $\omega_{01}^{+}$ and $\omega_{01}^{-}$. Apart from the effective low-pass response inherent to the digital demodulation (with a sampling rate of 100 $\mu$s) and acquisition chain, we do not apply additional filtering in post-processing.

To maximize state separation, each complex stream is rotated in the IQ plane such that the parity contrast lies predominantly along the real axis. After rotation, the two real-valued trajectories $s_{+}(t)$ and $s_{-}(t)$ encode the same underlying parity telegraph signal but inverted relative to each other. Compensating for that one signal is inverted we increase the parity-state SNR by forming a combined discriminator $x(t) = \frac{1}{2}[s_{+}(t) + s_{-}(t)]$. Using a gaussian mixture model 
$$
p(x)=w_0 \mathcal{N}\left(x ; \mu_0, \sigma_0\right)+w_1 \mathcal{N}\left(x ; \mu_1, \sigma_1\right),
$$
and computing the overlap between the two normalized components,
$$
\mathcal{O}=\int_{-\infty}^{\infty} \min \left[\mathcal{N}\left(x ; \mu_0, \sigma_0\right), \mathcal{N}\left(x ; \mu_1, \sigma_1\right)\right] d x
$$
we define the corresponding discrimination fidelity as $\mathcal{F} = 1-\mathcal{O}$ and post-select datasets with $\mathcal{F}>0.999$, corresponding to 37\% of all acquired data.

We extract parity switching events by thresholding the combined discriminator $x(t)$ to obtain a binary state record. We set the threshold at the midpoint between the two fitted Gaussian means, $\theta=\frac{\mu_0+\mu_1}{2}$,
$$
P(t)= \begin{cases}0, & x(t)<\theta, \\ 1, & x(t) \geq \theta .\end{cases}
$$
We identify switching events as transitions in $P(t)$.

\section{\label{app:D}Identification of burst intervals}
We identify burst intervals from the extracted parity-switching timestamps $\{x_i\}_{i=1}^{n}$, where $x_i$ denotes the occurrence time of the $i$th detected parity switch and $n$ is the number of events in the analyzed record. We first compute a smooth estimate of the event activity using kernel density estimation (KDE) with a Gaussian kernel,
\begin{equation}
\hat{f}_h(x)=\frac{1}{n h \sigma} \frac{1}{\sqrt{2 \pi}} \sum_{i=1}^n \exp \left(\frac{-\left(x-x_i\right)^2}{2 h^2 \sigma^2}\right),
\end{equation}
which serves as a continuous-time proxy for the switching density. Here $h$ is the kernel size and $\sigma$ the standard deviation of the samples. Elevated switching activity (bursts) manifests itself as pronounced local maxima in $\hat{f}_h(x)$. We use a size $h=20~\mathrm{ms}$ chosen to be larger than the typical burst duration and to reduce computational costs we compute $\hat{f}_h(x)$ in segments of $2~\mathrm{s}$.

Within each segment we identify local maxima of $\hat{f}_h(x)$ and apply an adaptive peak-height threshold. Denoting the set of peak heights within the segment as ${p_j}$ we define the threshold as
$$
p_{\mathrm{th}}=\mathrm{median}(\{p_j\})+2\,\mathrm{std}(\{p_j\})
%p_{\mathrm{th}}=\mathrm{median}({p_j})+2\std({p_j})
$$
and retain peaks satisfying $p_j>p_{\mathrm{th}}$ as candidate burst centers. Each retained peak is mapped to the nearest event timestamp index, yielding a set of candidate burst center indices ${i_c}$.

For each candidate center $i_c$, we look for events within a time window of $\pm 20~\mathrm{ms}$ around $x_{i_c}$ and form preliminary burst groups by requiring consecutive inter-event spacings of $x_{k+1}-x_k \le 5~\mathrm{ms}$. The final burst selection follows the main text definition: we retain only groups containing at least three events. We define the burst start and end times as the time stamps of the first and last events in the group, and the burst duration as $\tau_{\mathrm{burst}}=t_{\mathrm{end}}-t_{\mathrm{start}}$.

\section{Effective offset charge}
We convert the measured parity splitting into an effective offset charge coordinate. Because our spectroscopy yields the absolute parity splitting $\Delta f_{\mathrm{abs}} \equiv |\omega_{01}^{+}-\omega_{01}^{-}|/2\pi$, we use the expected charge-dispersion dependence for a charge-sensitive transmon, $\Delta f_{\mathrm{abs}}(n_g)\approx A|\cos(2\pi n_g)|$, where $A$ is the maximum frequency splitting. We estimate $A$ from the spectroscopy data and invert the relation pointwise to obtain a effective charge coordinate $n_g^{(\mathrm{eff})}(t)\in[0,0.25]$ via
$$
n_g^{(\text {eff})}(t)=\frac{1}{2 \pi} \arccos \left(\frac{\Delta f_{\mathrm{abs}}(t)}{A}\right) .
$$

%\bibliography{References_fromZotero_QPD-Bursts}

\begin{thebibliography}{10}
\expandafter\ifx\csname url\endcsname\relax
  \def\url#1{\texttt{#1}}\fi
\expandafter\ifx\csname urlprefix\endcsname\relax\def\urlprefix{URL }\fi
\providecommand{\bibinfo}[2]{#2}
\providecommand{\eprint}[2][]{\url{#2}}

\bibitem{glazman_bogoliubov_2021}
\bibinfo{author}{Glazman, L.} \& \bibinfo{author}{Catelani, G.}
\newblock \bibinfo{title}{Bogoliubov quasiparticles in superconducting qubits}.
\newblock \emph{\bibinfo{journal}{SciPost Physics Lecture Notes}}
  \bibinfo{pages}{031} (\bibinfo{year}{2021}).
\newblock \urlprefix\url{https://scipost.org/SciPostPhysLectNotes.31}.

\bibitem{aumentado_nonequilibrium_2004}
\bibinfo{author}{Aumentado, J.}, \bibinfo{author}{Keller, M.~W.},
  \bibinfo{author}{Martinis, J.~M.} \& \bibinfo{author}{Devoret, M.~H.}
\newblock \bibinfo{title}{Nonequilibrium {Quasiparticles} and \$2e\$
  {Periodicity} in {Single}-{Cooper}-{Pair} {Transistors}}.
\newblock \emph{\bibinfo{journal}{Physical Review Letters}}
  \textbf{\bibinfo{volume}{92}}, \bibinfo{pages}{066802}
  (\bibinfo{year}{2004}).
\newblock
  \urlprefix\url{https://link.aps.org/doi/10.1103/PhysRevLett.92.066802}.
\newblock \bibinfo{note}{Publisher: American Physical Society}.

\bibitem{sun_measurements_2012}
\bibinfo{author}{Sun, L.} \emph{et~al.}
\newblock \bibinfo{title}{Measurements of {Quasiparticle} {Tunneling}
  {Dynamics} in a {Band}-{Gap}-{Engineered} {Transmon} {Qubit}}.
\newblock \emph{\bibinfo{journal}{Physical Review Letters}}
  \textbf{\bibinfo{volume}{108}}, \bibinfo{pages}{230509}
  (\bibinfo{year}{2012}).
\newblock
  \urlprefix\url{https://link.aps.org/doi/10.1103/PhysRevLett.108.230509}.
\newblock \bibinfo{note}{Publisher: American Physical Society}.

\bibitem{serniak_hot_2018}
\bibinfo{author}{Serniak, K.} \emph{et~al.}
\newblock \bibinfo{title}{Hot {Nonequilibrium} {Quasiparticles} in {Transmon}
  {Qubits}}.
\newblock \emph{\bibinfo{journal}{Physical Review Letters}}
  \textbf{\bibinfo{volume}{121}}, \bibinfo{pages}{157701}
  (\bibinfo{year}{2018}).
\newblock
  \urlprefix\url{https://link.aps.org/doi/10.1103/PhysRevLett.121.157701}.
\newblock \bibinfo{note}{Publisher: American Physical Society}.

\bibitem{tuokkola_methods_2025}
\bibinfo{author}{Tuokkola, M.} \emph{et~al.}
\newblock \bibinfo{title}{Methods to achieve near-millisecond energy relaxation
  and dephasing times for a superconducting transmon qubit}.
\newblock \emph{\bibinfo{journal}{Nature Communications}}
  \textbf{\bibinfo{volume}{16}}, \bibinfo{pages}{5421} (\bibinfo{year}{2025}).
\newblock \urlprefix\url{https://www.nature.com/articles/s41467-025-61126-0}.
\newblock \bibinfo{note}{Publisher: Nature Publishing Group}.

\bibitem{bland_millisecond_2025}
\bibinfo{author}{Bland, M.~P.} \emph{et~al.}
\newblock \bibinfo{title}{Millisecond lifetimes and coherence times in {2D}
  transmon qubits}.
\newblock \emph{\bibinfo{journal}{Nature}} \bibinfo{pages}{1--6}
  (\bibinfo{year}{2025}).
\newblock \urlprefix\url{https://www.nature.com/articles/s41586-025-09687-4}.
\newblock \bibinfo{note}{Publisher: Nature Publishing Group}.

\bibitem{acharya_quantum_2025}
\bibinfo{author}{Acharya, R.} \emph{et~al.}
\newblock \bibinfo{title}{Quantum error correction below the surface code
  threshold}.
\newblock \emph{\bibinfo{journal}{Nature}} \textbf{\bibinfo{volume}{638}},
  \bibinfo{pages}{920--926} (\bibinfo{year}{2025}).
\newblock \urlprefix\url{https://www.nature.com/articles/s41586-024-08449-y}.
\newblock \bibinfo{note}{Publisher: Nature Publishing Group}.

\bibitem{mcewen_resolving_2022}
\bibinfo{author}{McEwen, M.} \emph{et~al.}
\newblock \bibinfo{title}{Resolving catastrophic error bursts from cosmic rays
  in large arrays of superconducting qubits}.
\newblock \emph{\bibinfo{journal}{Nature Physics}}
  \textbf{\bibinfo{volume}{18}}, \bibinfo{pages}{107--111}
  (\bibinfo{year}{2022}).
\newblock \urlprefix\url{https://www.nature.com/articles/s41567-021-01432-8}.
\newblock \bibinfo{note}{Publisher: Nature Publishing Group}.

\bibitem{mcewen_resisting_2024}
\bibinfo{author}{McEwen, M.} \emph{et~al.}
\newblock \bibinfo{title}{Resisting {High}-{Energy} {Impact} {Events} through
  {Gap} {Engineering} in {Superconducting} {Qubit} {Arrays}}.
\newblock \emph{\bibinfo{journal}{Physical Review Letters}}
  \textbf{\bibinfo{volume}{133}}, \bibinfo{pages}{240601}
  (\bibinfo{year}{2024}).
\newblock
  \urlprefix\url{https://link.aps.org/doi/10.1103/PhysRevLett.133.240601}.
\newblock \bibinfo{note}{Publisher: American Physical Society}.

\bibitem{kurilovich_correlated_2025}
\bibinfo{author}{Kurilovich, V.~D.} \emph{et~al.}
\newblock \bibinfo{title}{Correlated {Error} {Bursts} in a {Gap}-{Engineered}
  {Superconducting} {Qubit} {Array}} (\bibinfo{year}{2025}).
\newblock \urlprefix\url{http://arxiv.org/abs/2506.18228}.
\newblock \bibinfo{note}{ArXiv:2506.18228 [quant-ph]}.

\bibitem{fowler_surface_2012}
\bibinfo{author}{Fowler, A.~G.}, \bibinfo{author}{Mariantoni, M.},
  \bibinfo{author}{Martinis, J.~M.} \& \bibinfo{author}{Cleland, A.~N.}
\newblock \bibinfo{title}{Surface codes: {Towards} practical large-scale
  quantum computation}.
\newblock \emph{\bibinfo{journal}{Physical Review A}}
  \textbf{\bibinfo{volume}{86}}, \bibinfo{pages}{032324}
  (\bibinfo{year}{2012}).
\newblock \urlprefix\url{https://link.aps.org/doi/10.1103/PhysRevA.86.032324}.
\newblock \bibinfo{note}{Publisher: American Physical Society}.

\bibitem{houzet_photon-assisted_2019}
\bibinfo{author}{Houzet, M.}, \bibinfo{author}{Serniak, K.},
  \bibinfo{author}{Catelani, G.}, \bibinfo{author}{Devoret, M.} \&
  \bibinfo{author}{Glazman, L.}
\newblock \bibinfo{title}{Photon-{Assisted} {Charge}-{Parity} {Jumps} in a
  {Superconducting} {Qubit}}.
\newblock \emph{\bibinfo{journal}{Physical Review Letters}}
  \textbf{\bibinfo{volume}{123}}, \bibinfo{pages}{107704}
  (\bibinfo{year}{2019}).
\newblock
  \urlprefix\url{https://link.aps.org/doi/10.1103/PhysRevLett.123.107704}.
\newblock \bibinfo{note}{Publisher: American Physical Society}.

\bibitem{yelton_correlated_2025}
\bibinfo{author}{Yelton, E.}, \bibinfo{author}{Larson, C.~P.},
  \bibinfo{author}{Dodge, K.}, \bibinfo{author}{Okubo, K.} \&
  \bibinfo{author}{Plourde, B. L.~T.}
\newblock \bibinfo{title}{Correlated quasiparticle poisoning from phonon-only
  events in superconducting qubits} (\bibinfo{year}{2025}).
\newblock \urlprefix\url{http://arxiv.org/abs/2503.09554}.
\newblock \bibinfo{note}{ArXiv:2503.09554 [quant-ph]}.

\bibitem{serniak_direct_2019}
\bibinfo{author}{Serniak, K.} \emph{et~al.}
\newblock \bibinfo{title}{Direct {Dispersive} {Monitoring} of {Charge} {Parity}
  in {Offset}-{Charge}-{Sensitive} {Transmons}}.
\newblock \emph{\bibinfo{journal}{Physical Review Applied}}
  \textbf{\bibinfo{volume}{12}}, \bibinfo{pages}{014052}
  (\bibinfo{year}{2019}).
\newblock
  \urlprefix\url{https://link.aps.org/doi/10.1103/PhysRevApplied.12.014052}.
\newblock \bibinfo{note}{Publisher: American Physical Society}.

\bibitem{gordon_environmental_2022}
\bibinfo{author}{Gordon, R.~T.} \emph{et~al.}
\newblock \bibinfo{title}{Environmental radiation impact on lifetimes and
  quasiparticle tunneling rates of fixed-frequency transmon qubits}.
\newblock \emph{\bibinfo{journal}{Applied Physics Letters}}
  \textbf{\bibinfo{volume}{120}}, \bibinfo{pages}{074002}
  (\bibinfo{year}{2022}).
\newblock \urlprefix\url{https://doi.org/10.1063/5.0078785}.

\bibitem{rehammar_low-pass_2023}
\bibinfo{author}{Rehammar, R.} \& \bibinfo{author}{Gasparinetti, S.}
\newblock \bibinfo{title}{Low-{Pass} {Filter} {With} {Ultrawide} {Stopband} for
  {Quantum} {Computing} {Applications}}.
\newblock \emph{\bibinfo{journal}{IEEE Transactions on Microwave Theory and
  Techniques}} \textbf{\bibinfo{volume}{71}}, \bibinfo{pages}{3075--3080}
  (\bibinfo{year}{2023}).
\newblock \urlprefix\url{https://ieeexplore.ieee.org/document/10032269}.

\bibitem{connolly_coexistence_2024}
\bibinfo{author}{Connolly, T.} \emph{et~al.}
\newblock \bibinfo{title}{Coexistence of {Nonequilibrium} {Density} and
  {Equilibrium} {Energy} {Distribution} of {Quasiparticles} in a
  {Superconducting} {Qubit}}.
\newblock \emph{\bibinfo{journal}{Physical Review Letters}}
  \textbf{\bibinfo{volume}{132}}, \bibinfo{pages}{217001}
  (\bibinfo{year}{2024}).
\newblock
  \urlprefix\url{https://link.aps.org/doi/10.1103/PhysRevLett.132.217001}.
\newblock \bibinfo{note}{Publisher: American Physical Society}.

\bibitem{andersson_co-designed_2025}
\bibinfo{author}{Andersson, L.}, \bibinfo{author}{Olsson, B.},
  \bibinfo{author}{Gasparinetti, S.} \& \bibinfo{author}{Rehammar, R.}
\newblock \bibinfo{title}{Co-{Designed} {Reflective} and {Leaky}-{Waveguide}
  {Low}-{Pass} {Filter} for {Superconducting} {Circuits}}.
\newblock \emph{\bibinfo{journal}{IEEE Transactions on Microwave Theory and
  Techniques}} \bibinfo{pages}{1--7} (\bibinfo{year}{2025}).
\newblock \urlprefix\url{https://ieeexplore.ieee.org/document/11311711}.

\bibitem{vepsalainen_impact_2020}
\bibinfo{author}{Vepsäläinen, A.~P.} \emph{et~al.}
\newblock \bibinfo{title}{Impact of ionizing radiation on superconducting qubit
  coherence}.
\newblock \emph{\bibinfo{journal}{Nature}} \textbf{\bibinfo{volume}{584}},
  \bibinfo{pages}{551--556} (\bibinfo{year}{2020}).
\newblock \urlprefix\url{https://www.nature.com/articles/s41586-020-2619-8}.
\newblock \bibinfo{note}{Publisher: Nature Publishing Group}.

\bibitem{cardani_reducing_2021}
\bibinfo{author}{Cardani, L.} \emph{et~al.}
\newblock \bibinfo{title}{Reducing the impact of radioactivity on quantum
  circuits in a deep-underground facility}.
\newblock \emph{\bibinfo{journal}{Nature Communications}}
  \textbf{\bibinfo{volume}{12}}, \bibinfo{pages}{2733} (\bibinfo{year}{2021}).
\newblock \urlprefix\url{https://www.nature.com/articles/s41467-021-23032-z}.
\newblock \bibinfo{note}{Publisher: Nature Publishing Group}.

\bibitem{martinis_saving_2021}
\bibinfo{author}{Martinis, J.~M.}
\newblock \bibinfo{title}{Saving superconducting quantum processors from decay
  and correlated errors generated by gamma and cosmic rays}.
\newblock \emph{\bibinfo{journal}{npj Quantum Information}}
  \textbf{\bibinfo{volume}{7}}, \bibinfo{pages}{90} (\bibinfo{year}{2021}).
\newblock \urlprefix\url{https://www.nature.com/articles/s41534-021-00431-0}.
\newblock \bibinfo{note}{Publisher: Nature Publishing Group}.

\bibitem{iaia_phonon_2022}
\bibinfo{author}{Iaia, V.} \emph{et~al.}
\newblock \bibinfo{title}{Phonon downconversion to suppress correlated errors
  in superconducting qubits}.
\newblock \emph{\bibinfo{journal}{Nature Communications}}
  \textbf{\bibinfo{volume}{13}}, \bibinfo{pages}{6425} (\bibinfo{year}{2022}).
\newblock \urlprefix\url{https://www.nature.com/articles/s41467-022-33997-0}.
\newblock \bibinfo{note}{Publisher: Nature Publishing Group}.

\bibitem{xu_distributed_2022}
\bibinfo{author}{Xu, Q.} \emph{et~al.}
\newblock \bibinfo{title}{Distributed {Quantum} {Error} {Correction} for
  {Chip}-{Level} {Catastrophic} {Errors}}.
\newblock \emph{\bibinfo{journal}{Physical Review Letters}}
  \textbf{\bibinfo{volume}{129}}, \bibinfo{pages}{240502}
  (\bibinfo{year}{2022}).
\newblock
  \urlprefix\url{https://link.aps.org/doi/10.1103/PhysRevLett.129.240502}.

\bibitem{thorbeck_two-level-system_2023}
\bibinfo{author}{Thorbeck, T.}, \bibinfo{author}{Eddins, A.},
  \bibinfo{author}{Lauer, I.}, \bibinfo{author}{McClure, D.~T.} \&
  \bibinfo{author}{Carroll, M.}
\newblock \bibinfo{title}{Two-{Level}-{System} {Dynamics} in a
  {Superconducting} {Qubit} {Due} to {Background} {Ionizing} {Radiation}}.
\newblock \emph{\bibinfo{journal}{PRX Quantum}} \textbf{\bibinfo{volume}{4}},
  \bibinfo{pages}{020356} (\bibinfo{year}{2023}).
\newblock \urlprefix\url{https://link.aps.org/doi/10.1103/PRXQuantum.4.020356}.
\newblock \bibinfo{note}{Publisher: American Physical Society}.

\bibitem{fowler_spectroscopic_2024}
\bibinfo{author}{Fowler, J.~W.} \emph{et~al.}
\newblock \bibinfo{title}{Spectroscopic {Measurements} and {Models} of {Energy}
  {Deposition} in the {Substrate} of {Quantum} {Circuits} by {Natural}
  {Ionizing} {Radiation}}.
\newblock \emph{\bibinfo{journal}{PRX Quantum}} \textbf{\bibinfo{volume}{5}},
  \bibinfo{pages}{040323} (\bibinfo{year}{2024}).
\newblock \urlprefix\url{https://link.aps.org/doi/10.1103/PRXQuantum.5.040323}.
\newblock \bibinfo{note}{Publisher: American Physical Society}.

\bibitem{harrington_synchronous_2025}
\bibinfo{author}{Harrington, P.~M.} \emph{et~al.}
\newblock \bibinfo{title}{Synchronous detection of cosmic rays and correlated
  errors in superconducting qubit arrays}.
\newblock \emph{\bibinfo{journal}{Nature Communications}}
  \textbf{\bibinfo{volume}{16}}, \bibinfo{pages}{6428} (\bibinfo{year}{2025}).
\newblock \urlprefix\url{https://www.nature.com/articles/s41467-025-61385-x}.
\newblock \bibinfo{note}{Publisher: Nature Publishing Group}.

\bibitem{larson_quasiparticle_2025}
\bibinfo{author}{Larson, C.} \emph{et~al.}
\newblock \bibinfo{title}{Quasiparticle {Poisoning} of {Superconducting}
  {Qubits} with {Active} {Gamma} {Irradiation}}.
\newblock \emph{\bibinfo{journal}{PRX Quantum}} \textbf{\bibinfo{volume}{6}},
  \bibinfo{pages}{030339} (\bibinfo{year}{2025}).
\newblock \urlprefix\url{https://link.aps.org/doi/10.1103/2lyd-8swv}.
\newblock \bibinfo{note}{Publisher: American Physical Society}.

\bibitem{li_cosmic-ray-induced_2025}
\bibinfo{author}{Li, X.} \emph{et~al.}
\newblock \bibinfo{title}{Cosmic-ray-induced correlated errors in
  superconducting qubit array}.
\newblock \emph{\bibinfo{journal}{Nature Communications}}
  \textbf{\bibinfo{volume}{16}}, \bibinfo{pages}{4677} (\bibinfo{year}{2025}).
\newblock \urlprefix\url{https://www.nature.com/articles/s41467-025-59778-z}.
\newblock \bibinfo{note}{Publisher: Nature Publishing Group}.

\bibitem{valenti_spatial_2025}
\bibinfo{author}{Valenti, F.} \emph{et~al.}
\newblock \bibinfo{title}{The spatial correlation of radiation-induced errors
  in superconducting devices decays over a millimeter} (\bibinfo{year}{2025}).
\newblock \urlprefix\url{http://arxiv.org/abs/2505.04902}.
\newblock \bibinfo{note}{ArXiv:2505.04902 [cond-mat]}.

\bibitem{nho_recovery_2025}
\bibinfo{author}{Nho, H.} \emph{et~al.}
\newblock \bibinfo{title}{Recovery dynamics of a gap-engineered transmon after
  a quasiparticle burst} (\bibinfo{year}{2025}).
\newblock \urlprefix\url{http://arxiv.org/abs/2505.08104}.
\newblock \bibinfo{note}{ArXiv:2505.08104 [quant-ph]}.

\bibitem{wilen_correlated_2021}
\bibinfo{author}{Wilen, C.~D.} \emph{et~al.}
\newblock \bibinfo{title}{Correlated charge noise and relaxation errors in
  superconducting qubits}.
\newblock \emph{\bibinfo{journal}{Nature}} \textbf{\bibinfo{volume}{594}},
  \bibinfo{pages}{369--373} (\bibinfo{year}{2021}).
\newblock \urlprefix\url{https://www.nature.com/articles/s41586-021-03557-5}.
\newblock \bibinfo{note}{Publisher: Nature Publishing Group}.

\bibitem{christensen_anomalous_2019}
\bibinfo{author}{Christensen, B.~G.} \emph{et~al.}
\newblock \bibinfo{title}{Anomalous charge noise in superconducting qubits}.
\newblock \emph{\bibinfo{journal}{Physical Review B}}
  \textbf{\bibinfo{volume}{100}}, \bibinfo{pages}{140503}
  (\bibinfo{year}{2019}).
\newblock \urlprefix\url{https://link.aps.org/doi/10.1103/PhysRevB.100.140503}.
\newblock \bibinfo{note}{Publisher: American Physical Society}.

\bibitem{riste_millisecond_2013}
\bibinfo{author}{Ristè, D.} \emph{et~al.}
\newblock \bibinfo{title}{Millisecond charge-parity fluctuations and induced
  decoherence in a superconducting transmon qubit}.
\newblock \emph{\bibinfo{journal}{Nature Communications}}
  \textbf{\bibinfo{volume}{4}}, \bibinfo{pages}{1913} (\bibinfo{year}{2013}).
\newblock \urlprefix\url{https://www.nature.com/articles/ncomms2936}.
\newblock \bibinfo{note}{Publisher: Nature Publishing Group}.

\bibitem{amin_direct_2024}
\bibinfo{author}{Amin, K.~R.} \emph{et~al.}
\newblock \bibinfo{title}{Direct detection of quasiparticle tunneling with a
  charge-sensitive superconducting sensor coupled to a waveguide}
  (\bibinfo{year}{2024}).
\newblock \urlprefix\url{http://arxiv.org/abs/2404.01277}.
\newblock \bibinfo{note}{ArXiv:2404.01277 [quant-ph]}.

\bibitem{fink2024}
\bibinfo{author}{Fink, C.~W.}, \bibinfo{author}{Salemi, C.~P.},
  \bibinfo{author}{Young, B.~A.}, \bibinfo{author}{Schuster, D.~I.} \&
  \bibinfo{author}{Kurinsky, N.~A.}
\newblock \bibinfo{title}{The {Superconducting} {Quasiparticle}-{Amplifying}
  {Transmon}: {A} {Qubit}-{Based} {Sensor} for {meV} {Scale} {Phonons} and
  {Single} {THz} {Photons}} (\bibinfo{year}{2024}).
\newblock \urlprefix\url{http://arxiv.org/abs/2310.01345}.
\newblock \bibinfo{note}{ArXiv:2310.01345 [hep-ex, physics:physics,
  physics:quant-ph]}.

\bibitem{koch_charge-insensitive_2007}
\bibinfo{author}{Koch, J.} \emph{et~al.}
\newblock \bibinfo{title}{Charge-insensitive qubit design derived from the
  {Cooper} pair box}.
\newblock \emph{\bibinfo{journal}{Physical Review A}}
  \textbf{\bibinfo{volume}{76}}, \bibinfo{pages}{042319}
  (\bibinfo{year}{2007}).
\newblock \urlprefix\url{https://link.aps.org/doi/10.1103/PhysRevA.76.042319}.
\newblock \bibinfo{note}{Publisher: American Physical Society}.

\bibitem{krinner_engineering_2019}
\bibinfo{author}{Krinner, S.} \emph{et~al.}
\newblock \bibinfo{title}{Engineering cryogenic setups for 100-qubit scale
  superconducting circuit systems}.
\newblock \emph{\bibinfo{journal}{EPJ Quantum Technology}}
  \textbf{\bibinfo{volume}{6}}, \bibinfo{pages}{2} (\bibinfo{year}{2019}).
\newblock \urlprefix\url{https://doi.org/10.1140/epjqt/s40507-019-0072-0}.

\bibitem{diamond_distinguishing_2022}
\bibinfo{author}{Diamond, S.} \emph{et~al.}
\newblock \bibinfo{title}{Distinguishing {Parity}-{Switching} {Mechanisms} in a
  {Superconducting} {Qubit}}.
\newblock \emph{\bibinfo{journal}{PRX Quantum}} \textbf{\bibinfo{volume}{3}},
  \bibinfo{pages}{040304} (\bibinfo{year}{2022}).
\newblock \urlprefix\url{https://link.aps.org/doi/10.1103/PRXQuantum.3.040304}.
\newblock \bibinfo{note}{Publisher: American Physical Society}.

\bibitem{perelshtein2022}
\bibinfo{author}{Perelshtein, M.R.} \emph{et~al.}
\newblock \bibinfo{title}{Broadband {Continuous}-{Variable} {Entanglement} {Generation} {Using} a {Kerr}-{Free} {Josephson} {Metamaterial}}.
\newblock \emph{\bibinfo{journal}{Physical Review Applied}},
  \bibinfo{pages}{024063} (\bibinfo{year}{2022}).
\newblock \urlprefix\url{https://link.aps.org/doi/10.1103/PhysRevApplied.18.024063}.
\newblock \bibinfo{note}{Publisher: American Physical Society}.

\end{thebibliography}

\end{document}